\documentclass[twocolumn,amsmath,amssymb,floatfix,pre,unsortedaddress]{revtex4} 

\usepackage{graphicx,color}
\definecolor{brown}{rgb}{0.63,0.27,0.18}
\definecolor{orange}{rgb}{1.00,0.65,0.00}

\usepackage{dcolumn}
\usepackage{bm,multirow}

\begin{document}




\title{Merging 1D and 3D genomic information: Challenges in modelling and validation}

\author{Alessandra Merlotti}
\email{alessandra.merlotti2@unibo.it}
\affiliation{Department of Physics and Astronomy (DIFA), University of Bologna, Viale Berti Pichat 6/2, 40127 Bologna Italy, and INFN Sez. Bologna, Italy}
\author{Angelo Rosa}
\email{anrosa@sissa.it}
\affiliation{SISSA - Scuola Internazionale Superiore di Studi Avanzati, Via Bonomea 265, 34136 Trieste, Italy}
\author{Daniel Remondini}
\email{daniel.remondini@unibo.it }
\affiliation{Department of Physics and Astronomy (DIFA), University of Bologna, Viale Berti Pichat 6/2, 40127 Bologna Italy, and INFN Sez. Bologna, Italy}



\begin{abstract}
Genome organization in eukaryotes 
during interphase 
stems from the delicate balance between
    non-random correlations present in the DNA polynucleotide linear sequence
    and
    the physico/chemical reactions which shape continuously the form and structure of DNA and chromatin inside the nucleus of the cell.
    It is now clear that these mechanisms have a key role in important processes like gene regulation,
    yet the detailed ways they act simultaneously and, eventually, come to influence each other even across very different length-scales remain largely unexplored.
    In this paper, we recapitulate some of the main results concerning gene regulatory and physical mechanisms, in relation to the information encoded in the 1D sequence and the 3D folding structure of DNA.
    In particular, we stress how reciprocal crossfeeding between 1D and 3D models may provide original insight into how these complex processes work and influence each other.
\end{abstract}

\maketitle

\begin{figure*}
    \centering
    \includegraphics[width=1\textwidth]{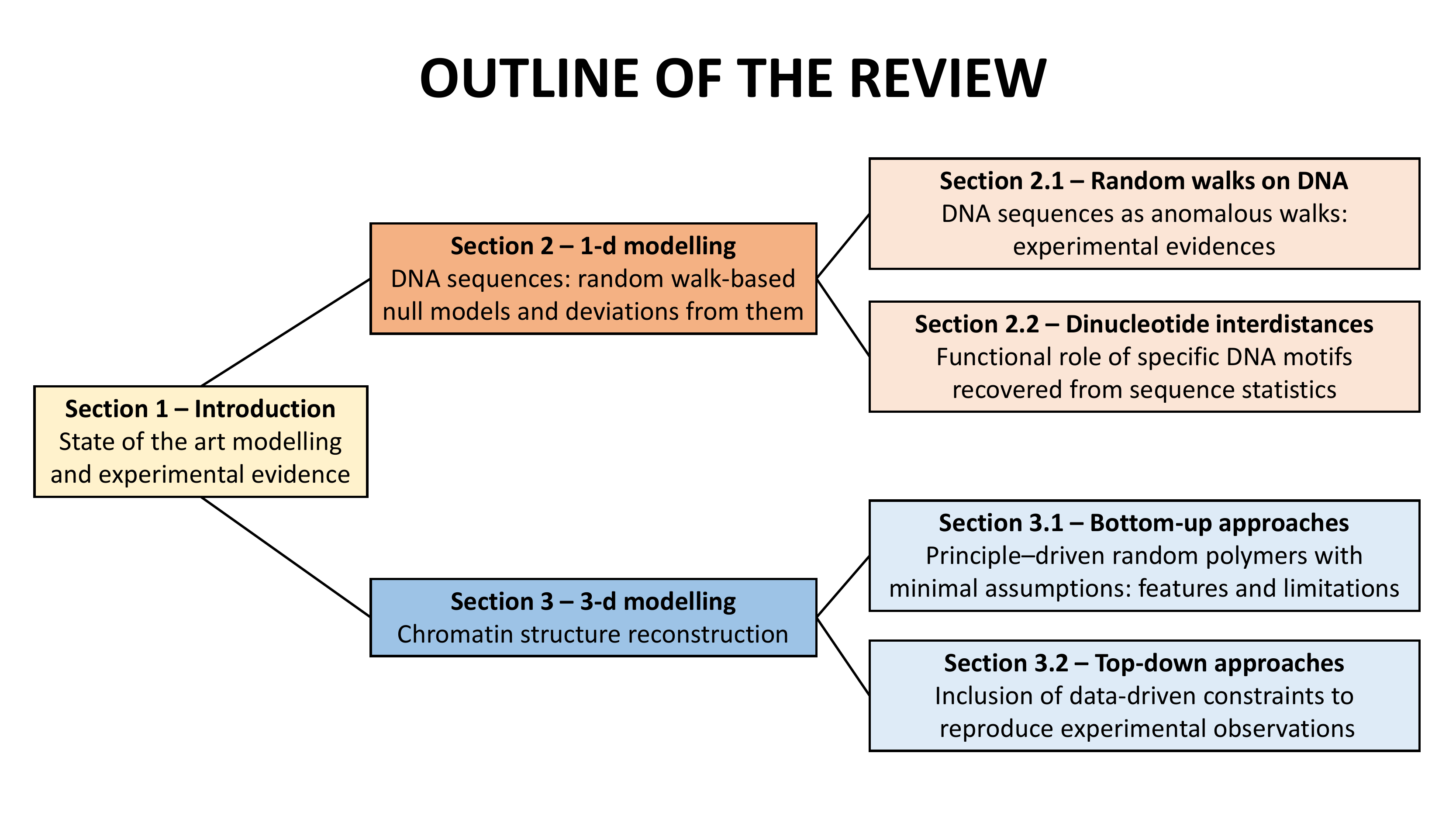}
    \label{fig:flowchart}
\end{figure*}

\section{Introduction}\label{sec:Intro}
\noindent

The interplay between the 1D sequence and the 3D folding of DNA in chromosomes and cell nuclei is mediated by the delicate balance between classical physical forces stemming from the DNA nature as a long, tightly packed polymer filament~\cite{GrosbergPolymSciReview2012,HalversonGrosbergReview2014,RosaZimmerReview2014,BiancoNicodemiReview2017,JostVaillantMeisterReview2017,JostVaillantRosaEveraersReview2017} and complex chemical processes governing DNA and histone methylations, nucleosome positioning and the binding of transcription factors to DNA sequence~\cite{Arneodo2011,Cortini2016} whose actions represent fundamental driving mechanisms in cell-fate decision~\cite{Stadhouders2019}.
For these reasons, understanding how the 1D genome affects its 3D spatial organization (and, viceversa) is a challenging task that requires a deeper understanding of both,
the physico/chemical forces governing DNA folding and the mechanisms beyond gene regulation:
advancing along this ambitious direction is compelling now more than ever, as it stands as the prerequisite for the comprehension of complex pathologies such as cancer \cite{Rippe2019}, laminopathies and premature aging diseases like Hutchinson-Gilford progeria and Werner syndromes \cite{Heyn2013,Dahl2006}.

\begin{figure}
    \centering
    \includegraphics[width=0.48\textwidth]{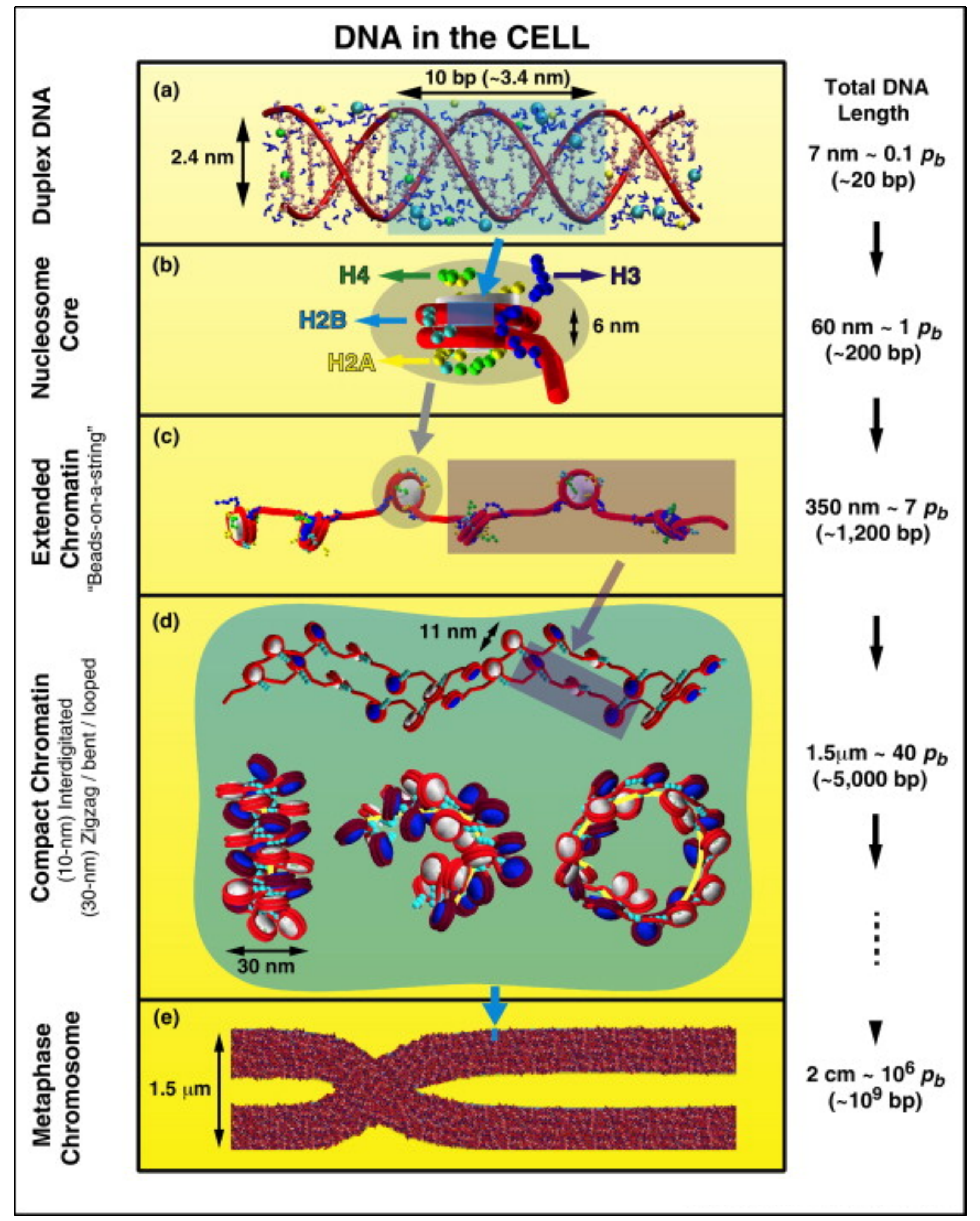}
    \caption{
    {\small
    Principles of chromosome folding. I.
    Schemating cartoon of the 10nm-fiber structure resulting from DNA wrapping around the histone complex.
    Chromatin folding beyond the 10nm-fiber up to the scale of the whole chromosome remains controversial.
    Reproduced with permission from Ref.~\cite{OzerSchlick-COSB2015}.
    }
    }
    \label{fig:ChromosomeFolding-FISH-HiC-pt1}
\end{figure}

In eukaryotes, 
every $\approx 200$ basepairs of the long DNA filament of each chromosome wrap around the histone complex~\cite{AlbertsBook}, by creating a necklace-like linear sequence of {\it nucleosomes}, 
commonly known as the 10nm chromatin fiber, see Fig.~\ref{fig:ChromosomeFolding-FISH-HiC-pt1}.
The present understanding of chromosome organization on spatial scales beyond the 10nm-fiber (in particular with respect to the existence of the ``elusive'' 30nm-fiber~\cite{Tremethick30nmFiberReview2007,OzerSchlick-COSB2015,NishinoMaeshimaEMBOJ2012,MaeshimaEMBOJ2016,MaeshimaLiquidChromatinCOGD2016,OSheaChromEMTScience2017}) appears still remarkably confused. 

However ambitious though, merging the information coming from the 1D/3D levels of knowledge promises to be increasingly affordable in the next future especially thanks to the recent, dramatic progress in sequencing techniques, such as the recent ATAC-seq and ChIA-Drop, which helped gaining new insights into the comprehension of 3D DNA organization as a function of 1D epigenetic ``marks'', in particular by allowing to map chromatin accessibility and nucleosome positioning genome-wide in a faster and more sensitive way than MNase-seq and DNase-seq \cite{Buenrostro2015} as for ATAC-seq, and revealing promoter-centred multivalent interactions in the ChIA-Drop case~\cite{Zheng2019}.

\begin{figure}
    \centering
    \includegraphics[width=0.48\textwidth]{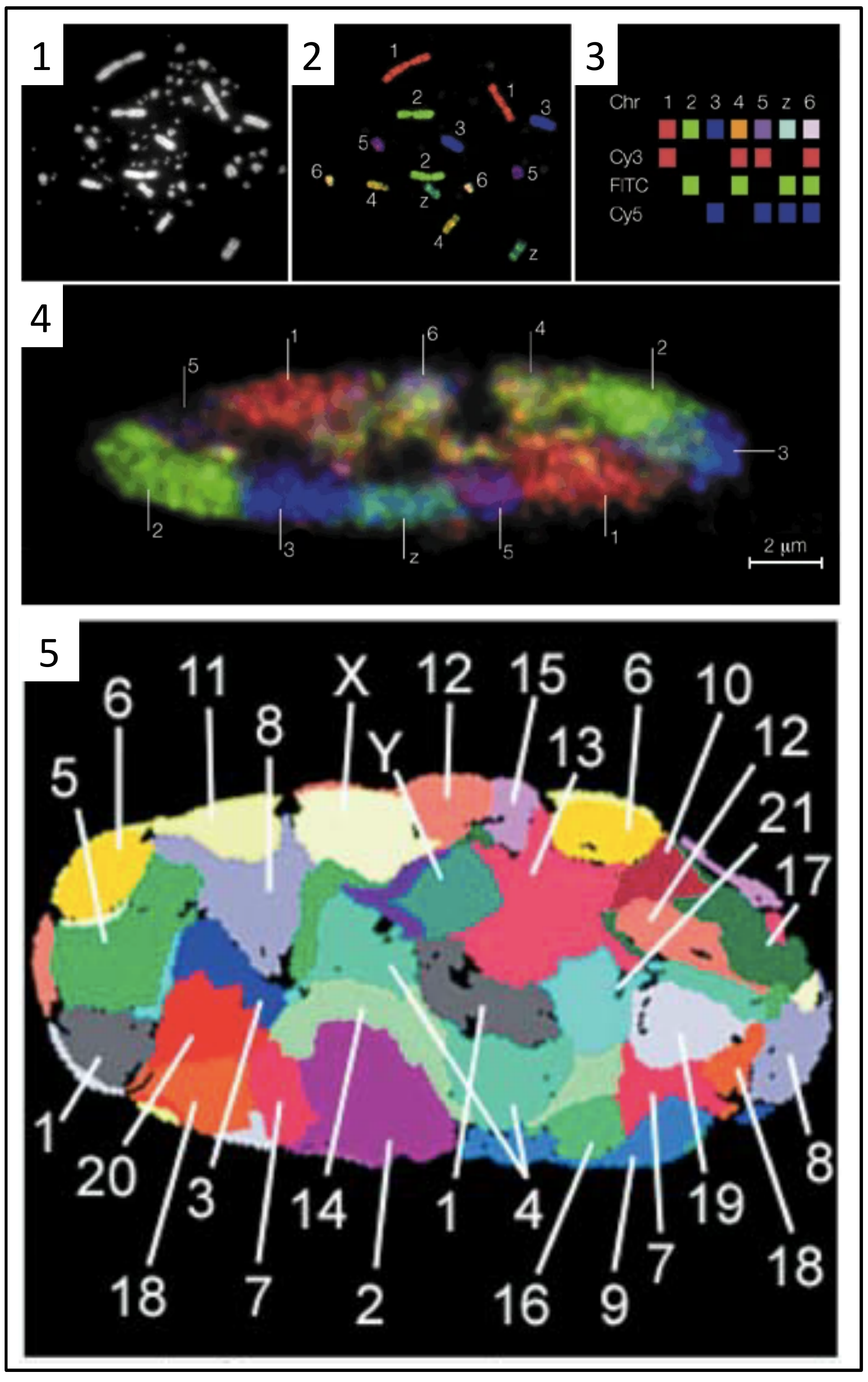}
    \caption{
    {\small
    Principles of chromosome folding. II.
    Chromosome ``painting'' by FISH (panels 1 to 3) reveals that chromosomes occupy distinct territories within the nucleus:
    panel 4 and panel 5 show examples of chromosome territories in chicken and human fibroblasts, respectively.
    Panels 1 to 4 are reproduced with permission from Ref.~\cite{CremerBrosReview2001},
    Panel 5 is reproduced from Ref.~\cite{BolzerCremerBrosPlosBiol2005} under Creative Commons License.
    }
    }
    \label{fig:ChromosomeFolding-FISH-HiC-pt2}
\end{figure}
\begin{figure}
    \centering
    \includegraphics[width=0.47\textwidth]{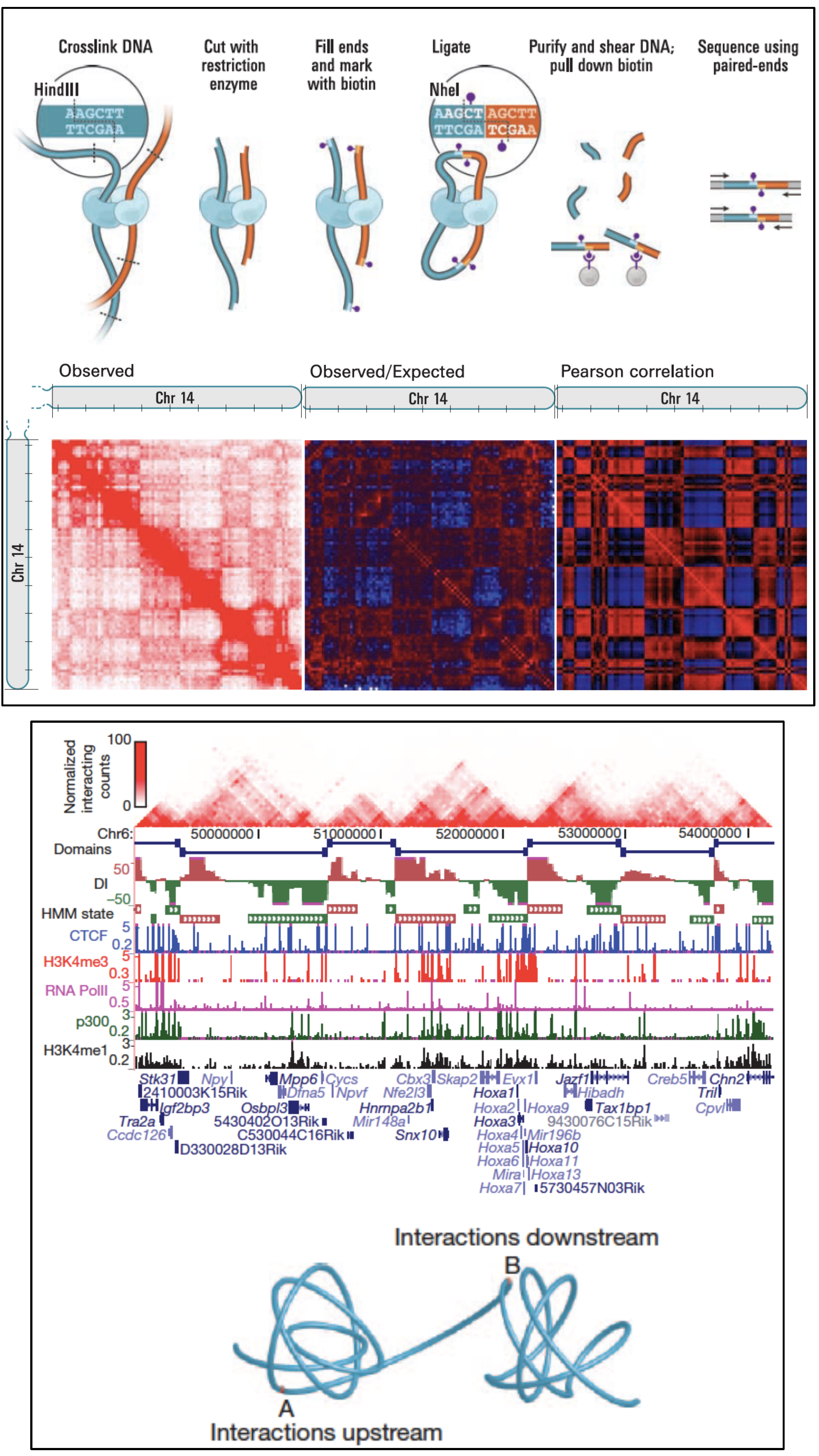}
    \caption{
    {\small
    Principles of chromosome folding. III.
    (Top)
    Chromatin contacts by HiC (top) at 1Mbp-resolution can be visualized in terms of heat maps.
    These maps show a ``plaid-pattern'' structure of intra-chromosome contacts stemming from chromosome compartmentalization into two (A/B) sub-compartments (bottom).
    Reproduced with permission from Ref.~\cite{LiebermanAiden-HiC-Science2009}).
    (Bottom)
    At higher resolution ($\lesssim 100$kbp),
    chromosomes appear organized into {\it topologically-associated domains} (TADs), regions characterized by unusually frequent contacts well separated by narrower regions almost interaction-depleted~\cite{DixonRenTADsNature2012,DixonRenTADsReview2016}.
    TADs correlate well with known epigenetic marks.
    Reproduced with permission from Ref.~\cite{DixonRenTADsNature2012}.
    }
    }
    \label{fig:ChromosomeFolding-FISH-HiC-pt3}
\end{figure}

At the same time, high-precision/high-resolution experimental techniques have now greatly contributed to expand our understanding of the physico/chemical properties of DNA {\it in vivo}:
\begin{itemize}
\item 
Chemical ``painting'' of DNA sequences by ``fluorescence {\it in situ} hibridization'' (FISH) (Fig.~\ref{fig:ChromosomeFolding-FISH-HiC-pt2}) shows that 
chromosomes fold into compact conformations 
(chromosome ``territories''~\cite{CremerBrosReview2001,CremerBrosReview2010}),
which have non-random, gene-correlated locations inside the nucleus~\cite{BolzerCremerBrosPlosBiol2005} and are crucial to cell correct behavior~\cite{CremerBrosReview2001,BolzerCremerBrosPlosBiol2005,CremerBrosReview2010}: 
in particular, territories help keeping some sort of ``physical barrier'' between close-by chromosomes (see Fig.~\ref{fig:ChromosomeFolding-FISH-HiC-pt2}),
with minimal amount of tangling~\cite{BrancoPomboPlosBiol2006} at the borders.
\item
Then, the internal structure inside each territory discloses itself by {\it chromosome conformation capture} techniques (3C~\cite{Dekker3CScience2002}) and HiC~\cite{LiebermanAiden-HiC-Science2009}), 
which are based on chromatin-chromatin cross-linking followed by DNA sequencing 
(Fig.~\ref{fig:ChromosomeFolding-FISH-HiC-pt3}, top):
this procedure showed that chromosomes display a checkerboard pattern of interactions~\cite{LiebermanAiden-HiC-Science2009}
revealing some compartmentalization into open/closed mega-basepair-sized sub-domains (Fig.~\ref{fig:ChromosomeFolding-FISH-HiC-pt3}, top).
At higher resolution, chromosomes cluster~\cite{DixonRenTADsNature2012} into ``topologically-associated domains'' (TADs),
regions separated by boundaries enriched for specific protein factors and identified by the unusually high number of contacts recorded in the each TAD's interior which drops suddenly at the boundaries (see the heat maps in Fig.~\ref{fig:ChromosomeFolding-FISH-HiC-pt3}, bottom).
Interestingly, chromosome organization into TADs appears ``universal'', being both stable across different cell lines and across different species~\cite{DixonRenTADsReview2016}.
\end{itemize}

With these premises, the hard core of the challenge lies in elaborating appropriate models that take into account or even integrate the 1D/3D levels of information.
In this review, we discuss the state-of-the-art of computational approaches which -- in our opinion -- have best contributed to shed new light on this fascinating and promising research field.
To this purpose, we adopt the following outline:
\begin{itemize}
    \item 
    In Section~\ref{sec:1dModels} we discuss 1D models for understanding non-random features in DNA sequences.
    \item
    In Sec.~\ref{sec:3dModelsChrOrg} we present a comprehensive catalogue of theoretical models based on polymer physics which describe relevant aspects of chromosome folding.
    Unless stated differently, we consider models for chromosome conformations during interphase~\cite{AlbertsBook}, {\it i.e} chromosomes within nuclear confinement.
\end{itemize}
In both sections we talk mainly of genome structure in higher eukaryotes (like mammals),
occasionally though we generalize to other classes of organisms.
Finally, we conclude the work (Sec.~\ref{sec:Disc}) by highlighting promising directions for future work, in particular with regard to what one can possibly learn by exploiting the connections between 1D and 3D modeling approaches.


%
\section{Reading the sequence: 1D models for nucleotide organization}\label{sec:1dModels}
At the 1D level, the DNA sequence can be represented as an ordinary string of text composed by four letters corresponding to the four nucleotides: A, C, G, T. This simple representation allowed to treat genomes as symbolic sequences and thus to exploit the knowledge developed in the fields of physics and statistics to extract information about their structure. In particular, two approaches have revealed helpful to identify some peculiar structural properties of genomic sequences that are involved in gene expression regulation, such as coding and non-coding regions~\cite{Peng1992,Peng1994}, enhancers \cite{Singh2018} and CpG islands \cite{Afreixo2015}: DNA random walks and dinucleotide interdistance.

\subsection{Random walks on DNA sequences}\label{sec:RWsDNAseqs}
\begin{figure}
    \centering
    \includegraphics[width=0.5\textwidth]{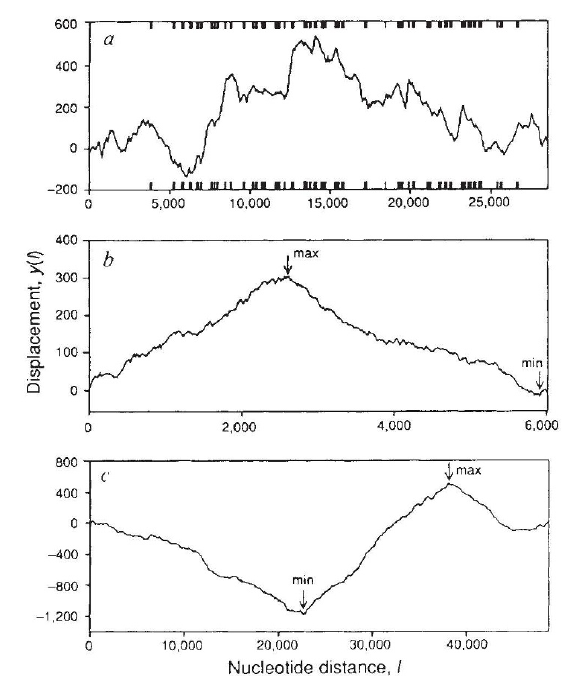}
    \caption{The DNA walk representation of intron-rich human $\beta$-cardiac myosin heavy-chain gene sequence (a), its cDNA (b), and the intron-less bacteriophage $\lambda$ DNA sequence (c). Note the more complex fluctuations for the intron-containing gene in (a)  compared with the intron-less sequences in (b) and (c). Heavy bars denote coding regions of the gene.
    Reproduced with permission from Ref.~\cite{Peng1992}.
    }
    \label{fig:RandomWalk_Peng}
\end{figure}

One of the first models of random walk on DNA sequence~\cite{Peng1992} was defined according to the following rule:
the walker steps up $(u(i) = +1)$ if a pyrimidine (`C' or `T' nucleotides) occurs at position $i$ along the sequence,
otherwise for the opposite case of a purine (`A' or `G' nucleotides) the walker steps down $(u(i) = -1)$.
This simple rule allows to calculate the displacement of a walker after $l$ steps as
\begin{equation}
    y(l) = \sum_{i=1}^{l} u(i)
\end{equation}
and to identify regions with different purine-pyrimidine content by plotting $y(l)$ as a function of nucleotide distance $l$ (see Fig.~\ref{fig:RandomWalk_Peng}), where positive slopes correspond to high concentration of pyrimidine and negative slopes correspond to high concentration of purines \cite{FractalBioBook}. 
The power of this simple approach is that different hypotheses on DNA sequence organization can be mapped onto specific ``null models'' about the characteristics of such random walks, and can thus be tested against the properties of the real sequences.
A fundamental statistical quantity characterizing any walk is the root mean square fluctuation $F(l)$ around the average displacement:
\begin{equation}
    F^{2}(l) = \overline{[\Delta y(l)]^{2}} - \overline{[\Delta y(l)]}^{2} 
\end{equation}
where $\Delta y(l) = y(l_{0} + l) - y(l_{0})$ and the bars indicate an average over all positions $l_{0}$ on the gene. 
The calculation of $F(l)$ is a key step in order to identify ``anomalous'' diffusion.
In fact, in pure ``random'' walks $F(l) \sim l^{1/2}$; otherwise, $F(l) \sim l^{\alpha}$, with $\alpha \neq 1/2$, thus revealing long-range correlations between walk steps, corresponding to correlations in nucleotide positioning process.
One of the earliest and most relevant results obtained by applying this method concerns the identification of coding and non-coding sequences inside genes~\cite{Peng1992}.
In particular, long-range correlations were identified as systematic markers of the presence of intron-containing genes and non-transcribed genomic regulatory elements, whereas, the absence of long-range correlations is characteristic of cDNA sequences and genes without introns (Fig.~\ref{fig:RandomWalk_Peng}).

Moreover,
long sequences (thousands of base pairs) were found inside non-coding regions, which were characterized by long-range correlations, and this led Buldyrev {\it et al.} \cite{Buldyrev1993} to apply a generalized L\'evy-walk model to non-coding sequences, and to hypothesize the existence of DNA loops. 
In generalized L\'evy-walks the typical walk step $l_{j}$ can be very long (in fact, walk steps are distributed according to a power-law distribution
$P(l_{j}) \propto 1/l_{j}^{\mu}$ with $2 < \mu < 3$), implying the existence of correlations between the displacements of nucleotides at very long mutual distances.

The authors provided a molecular basis for the power-law distribution of step lengths  by hypothesizing that, in order to be inserted into DNA, a macromolecule should form a loop of length $l_{j}$, whose ends come close to each other in the space. In fact, Buldyrev {\it et al.} pointed out that the long uncorrelated subsequences inside non-coding regions may correspond to repetitive elements, such as LINE-1, or retroviral sequences.


%
\subsection{Dinucleotide interdistance}\label{sec:DinuclInterdists}
Another approach results very powerful at identifying structural genomic features at the 1D level: the study of dinucleotide interdistance distributions.
The idea is inspired to the theory of first-return-time distributions in stochastic and deterministic processes by H. Poincar\'e, who developed this model to study the trajectories of bounded dynamical systems \cite{Poincare1890}. 

Referring to genome sequences, the analysis can be carried out through the following steps:
given a dinucleotide $XY$, where $X$ and $Y$ can take any value among $\{A,C,G,T\}$, its interdistance distribution $\hat{p}(\tau)$ can be calculated by
(i)
identifying the positions $x_{j}$ ($j=1, 2, ...$) of each $XY$ along the sequence,
(ii)
calculating the distance between two consecutive XY as $\tau_j \equiv x_{j+1} - x_{j}$,
(iii) counting the abundance of a given interdistance value $\tau$
and
(iv)
estimating its relative frequency $\hat{p}(\tau)$ according to the formula: 
\begin{equation}
    \hat{p}(\tau) =
    \frac{ \# \{ j=1, 2, ... | \tau_{j}=\tau \} }{ \# \{ j=1, 2, ... | \tau_{j} \} } \, ,
    \label{eq:interdist}
\end{equation}
where
the numerator counts all values where $\tau_j=\tau$ while the denominator runs over all unrestricted values $\tau_j$.

\begin{figure*}
    \centering
    \includegraphics[width=\textwidth]{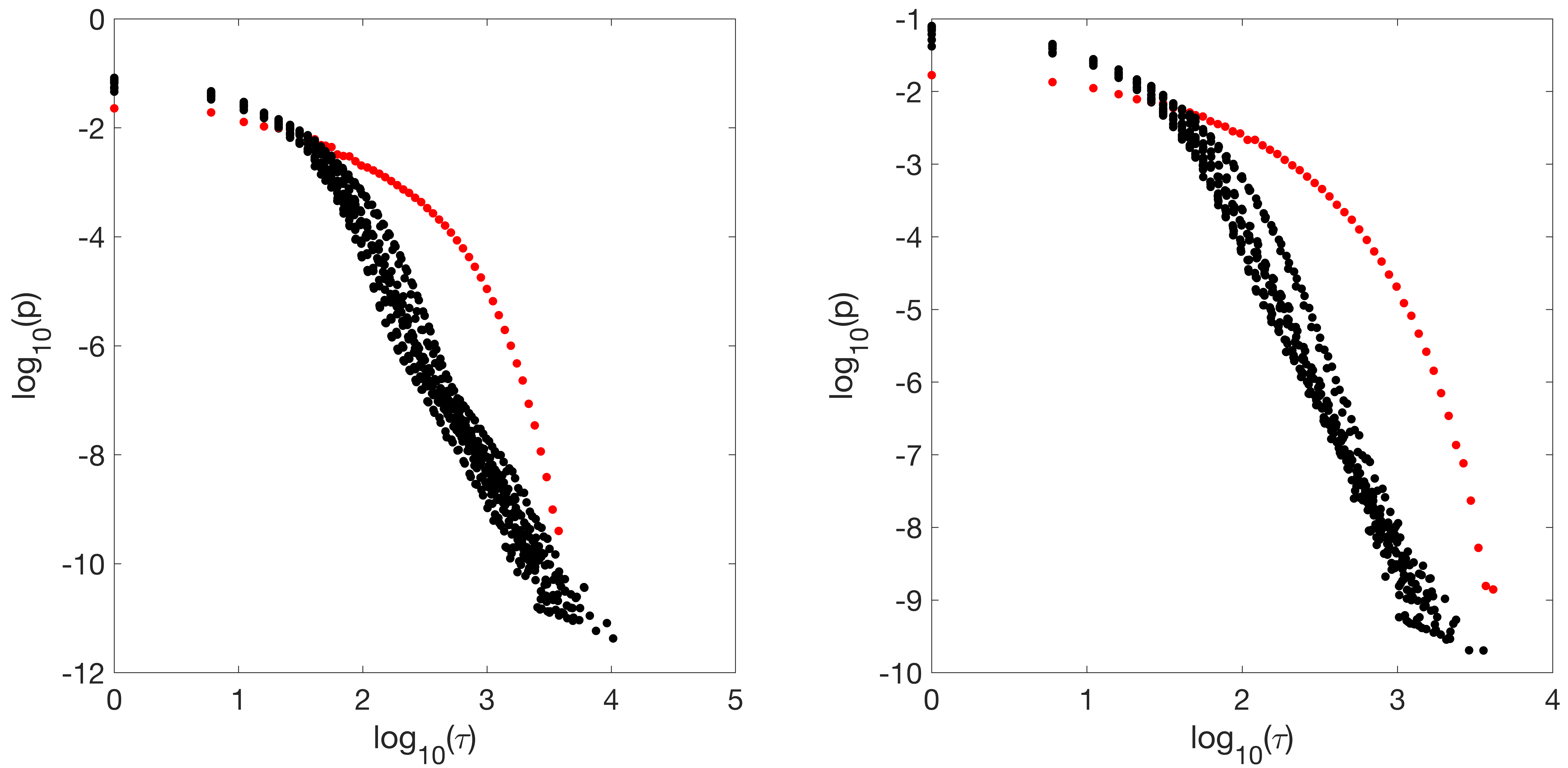} 
    \caption{
    Distribution functions of dinucleotide interdistances ($\tau$, measured in units of DNA basepairs) in log-log scale for {\it Homo sapiens} (left) and  {\it Mus musculus} (right).
    The distribution for CG dinucleotides is represented in red.
    }
    \label{fig:CG_Paci}
\end{figure*}

The first analysis of this quantity on genome sequences \cite{Afreixo2011} showed that dinucleotide interdistance distributions have a pronounced period-3 oscillatory behaviour in protein-coding regions which is absent in the whole-genome distributions and appears to be related to the triplet structure of the protein-coding genetic code.
Furthermore, the comparison between real distributions and randomly generated ones revealed that the behaviour of CG dinucleotides is considerably different from all the others. 
This study opened the avenue to subsequent works that led to methods for the identification of CpG islands \cite{Afreixo2015}, and to a more general characterization of CG interdistances in association to DNA methylation functionalities \cite{Paci2016,Merlotti2018}.
In particular, the work of Paci et al. \cite{Paci2016} revealed that CG interdistance distribution in higher-order organisms greatly differs from all other dinucleotides (see the comparison between
{\it Homo sapiens}
and 
{\it Mus musculus} 
in Fig.~\ref{fig:CG_Paci}), showing the strong exponential decay
\begin{equation}\label{eq:ExpDecay}
    \hat{p}(\tau) \sim e^{-\tau/b} \, .
\end{equation}
This difference seems to be related to the different role that methylation plays in this class of organisms \cite{Paci2016}. 
Interestingly, in higher-order organisms the characteristic ``length-scale'' $b$ of Eq.~(\ref{eq:ExpDecay}) measuring the average contour length distance between consecutive CGs showed a value $200 \mbox{ bp} < b < 300 \mbox{ bp}$, which is comparable to the typical DNA filament wrapped around the histone complex~\cite{AlbertsBook}.

\begin{figure*}
    \centering
    \includegraphics[width=1.0\textwidth]{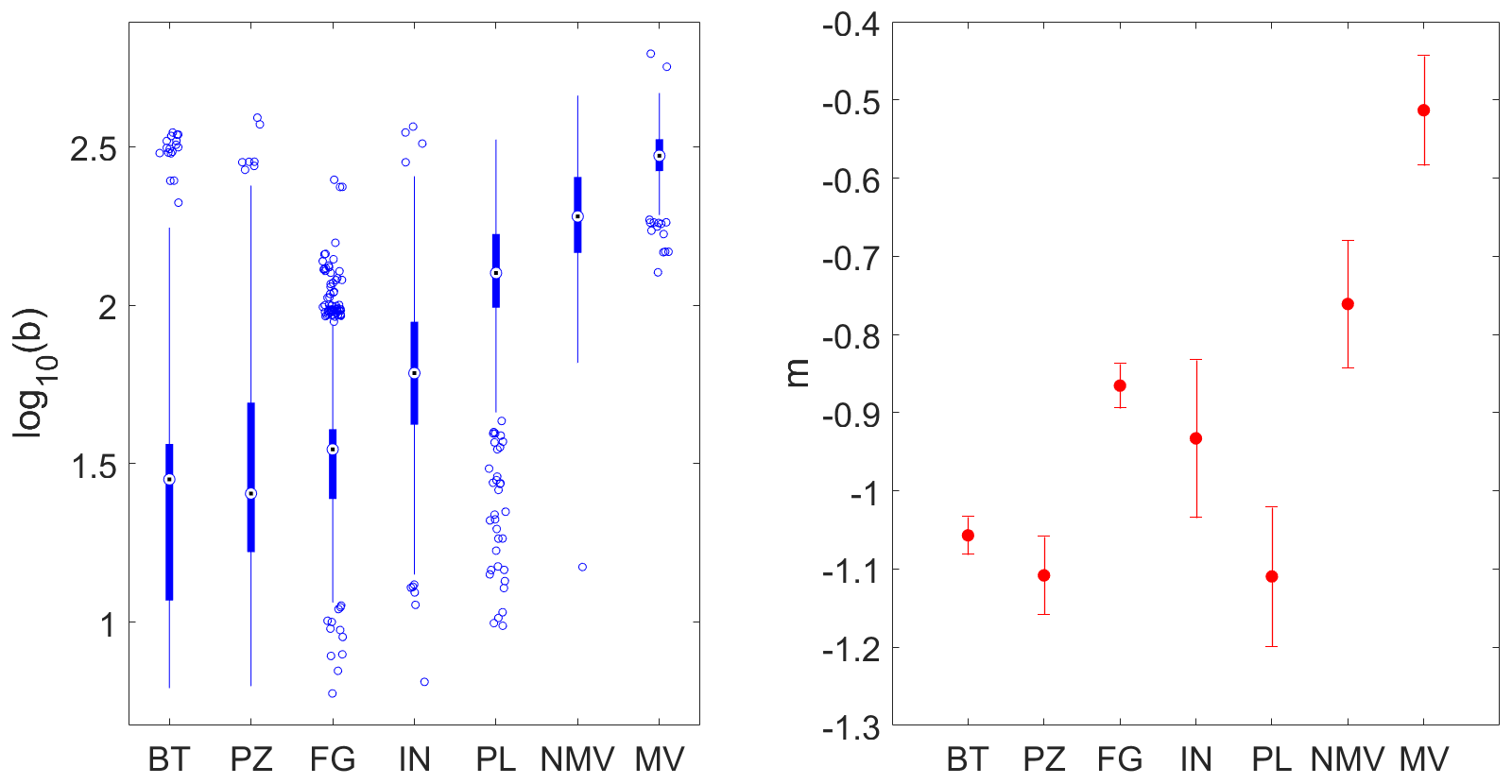}
    \caption{
    (Left)
    Box-plots for the Gamma-distribution scale parameters $b$ (see Eq.~(\ref{eq:GammaDistr})) for seven categories of organisms:
    bacteria (BT),
    protozoa (PZ),
    fungi (FG),
    invertebrates (IN),
    plants (PL),
    non-mammal vertebrates (NMV)
    and
    mammal vertebrates (MV).
    (Right)
    Estimated average values and error bars for $m$ exponents relative to the same classes of organisms.
    }
    \label{fig:CG_Merlotti}
\end{figure*}

An even deeper analysis of CG interdistance distributions was performed in human genome, by identifying the so-called Gamma-distribution 
\begin{equation}\label{eq:GammaDistr}
    \hat{p}(\tau) \sim \tau^{a-1} e^{-\tau/b}
\end{equation}
as the best fitting model distribution \cite{Merlotti2018}.
Furthermore, in this work the authors extended the study to a large variety of organisms spanning all available ranges of biological complexity, finding that the value of parameter $b$ is correlated to the biological complexity of the organism category:
in fact, it steadily increases moving from bacteria to vertebrates (see Fig.~\ref{fig:CG_Merlotti}, left)
and it is strongly correlated to CG density ($\mbox{CG}\%$),
displaying in particular a power-law behavior $b \propto \mbox{CG}\%^{m}$.
The study showed that all categories, except vertebrates, are characterized by an exponent $m \sim -1$, which is compatible with a simple null model predicting that the average distance between dinucleotides is inversely proportional to the dinucleotide density inside the sequence.
For vertebrates instead, the exponent $m$ takes the value $\simeq -0.5$
which is significantly higher in comparison to the other classes of organisms considered (see Fig.~\ref{fig:CG_Merlotti}, right):
we speculate that this might be related to a different mechanism for CG positioning along the genome connected to the DNA methylation process that CG dinucleotides undergo in this class of organisms.

These results show how detailed features of 1D DNA sequence 
are able to capture key properties of gene regulatory mechanisms that go beyond the 1D environment, such the extension of coding and noncoding regions, or the footprints of DNA methylation. 
In general, the relative positioning of specific DNA sequences along the genome might reflect their role in a specific 3D context,
in particular where complex loop structures can bring close to each other motifs quite far apart along the sequence, similar to what happens during the folding process of peptide chains.
The identification of the correct distributions of these distances can help to restrict the type of modelling processes able to generate them, thus helping to clarify the biology, chemistry and/or physics behind these far-from-trivial conformations.


%
\section{Folding the sequence: 3D models of chromosome organization in eukaryotes}\label{sec:3dModelsChrOrg}
``Predicting'' 3D chromosome structure starting from the 1D DNA sequence -- a question reminding in some way of the analogous protein folding problem~\cite{PandeGrosbergTanaka-RMP2000} -- is a long-standing problem in cell biology and a very challenging one.
Although the two problems (DNA folding and protein folding) may appear similar, a huge difference lies in the fact that DNA structure is not only guided by the chemical properties of its components (as for protein peptides) but relies on the complex interplay with many epigenetic factors (histones, noncoding RNAs, cohesins, lamines, etc.) that can be guided by ``signals'' set along the native DNA sequence (transcription factor binding sites, enhancer/promoter binding, DNA methylation, etc.) some of which, possibly, might be still unknown.
Moreover, chromosome state may depend on other important factors like, to mention a few, the particular cellular type, phase of the cellular cycle, gene activity and the mechanisms beyond DNA repair~\cite{AlbertsBook}.
Fortunately, the rapid development and increasing availability of structural data on chromosome organization (FISH and HiC {\it in primis}, see Sec.~\ref{sec:Intro}) alongside with the more and more sophisticate analysis tools (see Sec.~\ref{sec:1dModels}) which are now capable of detecting finer and finer correlations in the 1D DNA sequences are rapidly shifting the field towards a more confident description of how chromosomes fold inside the nucleus and how it reverberates on chromosome function.

As for it, in recent years there has been an impressive ``explosion'' of models trying to fill the missing conceptual gap between the 1D DNA or chromatin sequence and the 3D chromosome packing inside the nucleus.
Interestingly, most of (if not all) these models have been proposed by physicists and are based on the (rather obvious) assumption that chromosomes are long polymer chains subject to the same classical~\cite{DoiEdwardsBook,RubinsteinColbyBook} laws of polymer physics:
these laws can then be used to predict the {\it in vivo} chromosome behavior and, then, make quantitative and testable predictions.

As it has been stressed in the Introduction (Sec.~\ref{sec:Intro}), chromosome structure inside the nucleus remains highly controversial.
It is no surprise then, that there exists a conspicuous literature concerning different polymer models presenting alternative scenarios to illustrate the link between chromosome sequence and folding.
In the next sections we will discuss in more detail some of these models and the physical bases of each of them.

To better accomplish this purpose, it is instructive to classify the models into two categories:
\begin{enumerate}
    \item
    In the first category (Sec.~\ref{sec:BottomUpModels}), we place those models which rely on relatively few, minimal physical assumptions.
    The idea behind these approaches is that certain features of chromosome organization are common to all species and, in some respect, are {\it more important} than the details contained in each DNA sequence which make each species so different from any one else.
    Minimal models of this kind are extremely useful and instructive because they constitute the paradigm to understand the ``nuclear'' forces which continuously remodel the genomes.
    \item
    In the second category (Sec.~\ref{sec:TopDownModels}), we consider those polymer models which are constructed to satisfy a certain number of constraints obtained from experimental results.
    For this reason, we name these {\it data-driven  models}.
    These approaches are now becoming especially popular, for one hopes to employ them in the near future to provide accurate predictions on how genomes react when the ``native'' conditions upon which they were constructed change as the result of some stress on the cell or because of some induced mutation on the DNA sequence.
\end{enumerate}
%

%
\subsection{Chromosome organization by generic, ``bottom-up'' polymer physics}\label{sec:BottomUpModels}
\subsubsection{I. The role of topological constraints}\label{sec:TopologyRole}
Chromosomes are constituted by long chromatin filaments tightly packed inside the nucleus.
By neglecting all details related to the heterogeneity of DNA sequences,
at first approximation the entire system of chromosomes contained in the nucleus can be described as a solution of polymer chains~\cite{DoiEdwardsBook,RubinsteinColbyBook} subject to thermal fluctuations.
Under these conditions topological constraints, which are known to force nearby polymer chains to move randomly by {\it sliding} past each other without {\it passing} through each~\cite{DoiEdwardsBook,RubinsteinColbyBook}, are expected to play a key role by affecting chromosome structural and dynamical properties.

\begin{figure*}
    \centering
    \includegraphics[width=0.95\textwidth]{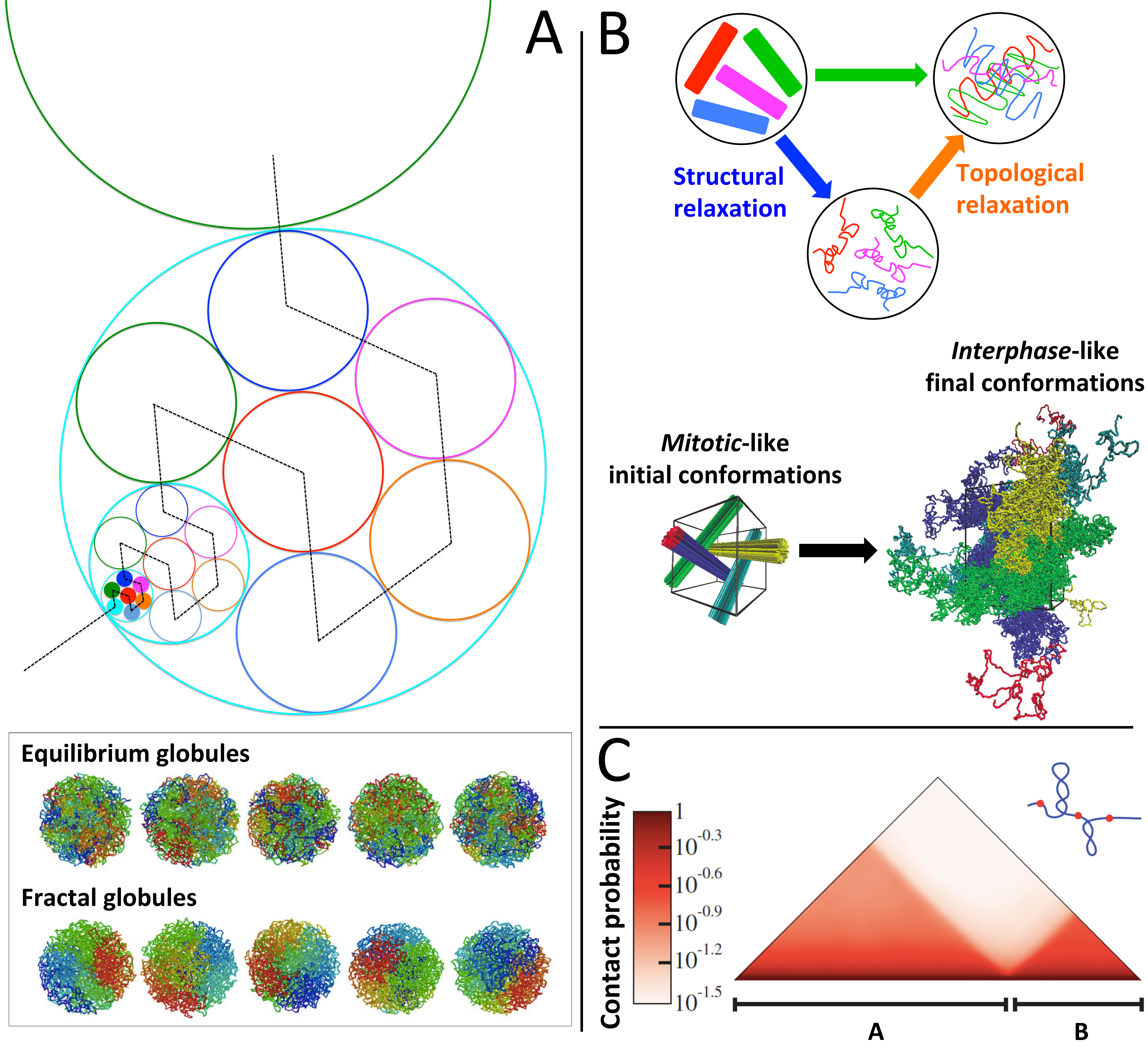}
    \caption{
    {\small
    The role of topological constraints in chromosome organization.
    (A, top)
    Schematic illustration of the ``crumpled globule'', showing the different layers in the hierarchical folding.
    The fundamental units (the monomers, filled spheres) fold into globular structures of larger sizes (the smaller empty spheres), acting in turn as ``super''-monomers in the following crumpling event. The process proceeds then at the next stage, and so on.
    The final structure resembles a fractal~\cite{GrosbergRabinEPL1993} with maximal compactness.
    (A, bottom)
    Examples of polymer conformations obtained by computer simulations, illustrating the structural differences between equilibrium and crumpled (fractal) globules.
    Reproduced with permission from Ref.~\cite{LiebermanAiden-HiC-Science2009}.
    (B)
    Because structural and topological relaxations of mitotic-like conformations have markedly different time-scales, chromosomes remain effectively ``trapped'' into territorial-like conformations~\cite{RosaEveraersPlos2008,RosaBeckerEveraersBJ2010,GrosbergPolymSciReview2012,HalversonGrosbergReview2014}.
    Reproduced from Ref.~\cite{RosaEveraersPlos2008} under Creative Commons License.
    (C)
    Chromatin fibers with (negative) levels of supercoiling form TAD-like structures~\cite{BenedettiStasiakNAR2013,BenedettiStasiakNAR2017}, reproducing contact patterns observed in HiC experiments.
    Reproduced from Ref.~\cite{BenedettiStasiakNAR2013} under Creative Commons CC BY License.
    }
    }
    \label{fig:TopologicalModelsSummary}
\end{figure*}

In fact, 
it is a non-trivial question to ask how a single centimeter-long chromosome can be efficiently stored inside the nucleus which is typically about thousand times~\cite{AlbertsBook} narrower.
While the presence of histone complexes and territories point towards the fact that chromosomes maintain a certain level of compactness,
they say nothing about how compactness can be practically and efficiently achieved.
In this respect, physical theories of polymers may become useful.

A major turning point occurred in the late '80s when Grosberg and colleagues published two influential papers~\cite{GrosbergNechaevShakhnovichJPhysFrance1988,GrosbergRabinEPL1993}, suggesting that the DNA or the chromatin fiber of a single chromosome should exist in an unknotted, off-equilibrium state which they termed ``the crumpled globule'', see Fig.~\ref{fig:TopologicalModelsSummary}(A).
Intuitively, this model can be constructed by assuming that the linear DNA sequence folds by hierarchical compaction from small up to the largest scales:
this fractal-like conformation features the two advantages of being maximally packed {\it and} knot-free.

From the theoretical point of view, two possible mechanisms leading to a crumpled globule were suggested:
either by fast switching the solvent conditions of the polymer chain from ``good'' to ``bad'' ({\it i.e.}, polymer self-interactions turn from repulsive to attractive~\cite{GrosbergNechaevShakhnovichJPhysFrance1988})
or by fast confinement of the polymer into a narrow region~\cite{LiebermanAiden-HiC-Science2009}.
Either way, the chain has no time to fully relax from its initial (knot-free) conformation,
the final state being crumpled and displaying the presence of domains.
Conversely, when the process of crumpling is slow, the final state is akin to an ``equilibrium'' globule with no domains (see the comparison between the two contrasting sets of model polymer conformations in Fig.~\ref{fig:TopologicalModelsSummary}(A)).
Although interesting from the theoretical point of view, fast crumpling is not expected to take place inside the cell.

In 1999, Langowski and collaborators introduced the so-called {\it random-loop model}~\cite{MunkelLangowskiJMB1999}:
interphase chromosome structure was described in terms of a self-repulsive random polymer with pairs of monomers permanently bound to form small loops on the scale of $\sim 100$kbp, rosette-like sub-compartments on larger scales and territories imposed by artificial confinement of the polymer chain.
The model was later~\cite{JhunjhunwalaCell2008} applied to describe the 3D structure of the murine immunoglobulin {\it heavy-chain} locus.
The random-loop model appears in qualitative agreement with chromatin organization into TADs and territories,
however this is not entirely surprising because these motifs were directly {\it imposed} on the model and, then, not really {\it explained}.

Instead, crumpled conformations can be easily obtained through a very simple physical mechanism which looks almost as the ``reverse'' of the one considered in the construction of a crumpled globule.
In two publications~\cite{RosaEveraersPlos2008,RosaBeckerEveraersBJ2010} Rosa and Everaers presented a polymer model for chromosome organization implying that territories emerge ``spontaneously'' as the result of the slow relaxation of the mitotic-like original chromosome structure (Fig.~\ref{fig:TopologicalModelsSummary}(B)):
in other words, the microscopic topological chromatin state remains quenched {\it in time} with no chance to relax and chromosomes get trapped into crumpled, territory-like conformations.
It was proposed~\cite{RosaEveraersPlos2008} then that the physical mechanism underlying chromosome compaction is the same driving the folding of untangled ring polymers in concentrated solutions~\cite{HalversonGrosbergReview2014,RosaEveraersPRL2014,SmrekKremerRosaACSML2019,SchramRosaEveraersSoftMatter2019}.
As demonstrated in~\cite{RosaEveraersPlos2008,RosaBeckerEveraersBJ2010}, the proposed model is able to capture {\it quantitatively} generic chromosome features like internal chromatin-chromatin distances and HiC contact frequencies with no fitting parameters, and can be used to model chromosome dynamics on time-scales from seconds to days in real time.
Third, it can be also naturally generalized~\cite{FlorescuTherizolsRosaPlosCB2016} so to take into account the heterogeneity of DNA sequence.

We conclude the section connecting chromosome organization and the topological properties of the chromatin fibers by mentioning some recent work by the Stasiak's group in Lausanne~\cite{BenedettiStasiakNAR2013,BenedettiStasiakNAR2017} which suggests a possible link between the presence of supercoiling in chromatin and TADs (mentioned in Sec.~\ref{sec:Intro}).
Chromosomal DNA is expected to be naturally supercoiled due to continuously ongoing processes like replication and transcription.
This excess of supercoiling is expected to never relax, once again because of the typically large size of chromosomes.
It may thus induce local crumpling of the chromatin fiber,
similar to what occurs to a familiar phone cord when excessive twist is applied.
By fine-tuning the amount of supercoiling in a numerical polymer model for chromatin fibers, Stasiak and colleagues showed that the phenomenology of TADs, summarized by the excess of intra-domain contacts with respect to inter-domain contacts (see Fig.~\ref{fig:TopologicalModelsSummary}(C)), can be generically captured.

\subsubsection{II. Sequence-specific chromatin-chromatin interactions}\label{sec:InteractionsRole}
The polymer models presented in Sec.~\ref{sec:TopologyRole} 
show that notable chromosome features like intra-DNA positions and contacts may be quantitatively understood in terms of the same theoretical mechanisms describing the phenomenology of entangled polymer solutions.
On the other hand, there is more to chromosome biology which requires a thorough discussion.

In this respect, it is known that certain species of protein complexes present in the nucleus tend to bind to specific DNA target sites and influence chromosome organization:
important examples include
the CCCTC binding factor (CTCF) involved in promoter activation or repression and 
methylation-dependent chromatin insulation~\cite{RendaCTCF2007}
and
the trascription units which by clustering into transcription ``factories''~\cite{CookTF2010} mediate and regulate the production of transcripts.
The role of these protein-DNA interactions in chromosome architecture has been addressed in an increasing number of publications.

\begin{figure}
    \centering
    \includegraphics[width=0.45\textwidth]{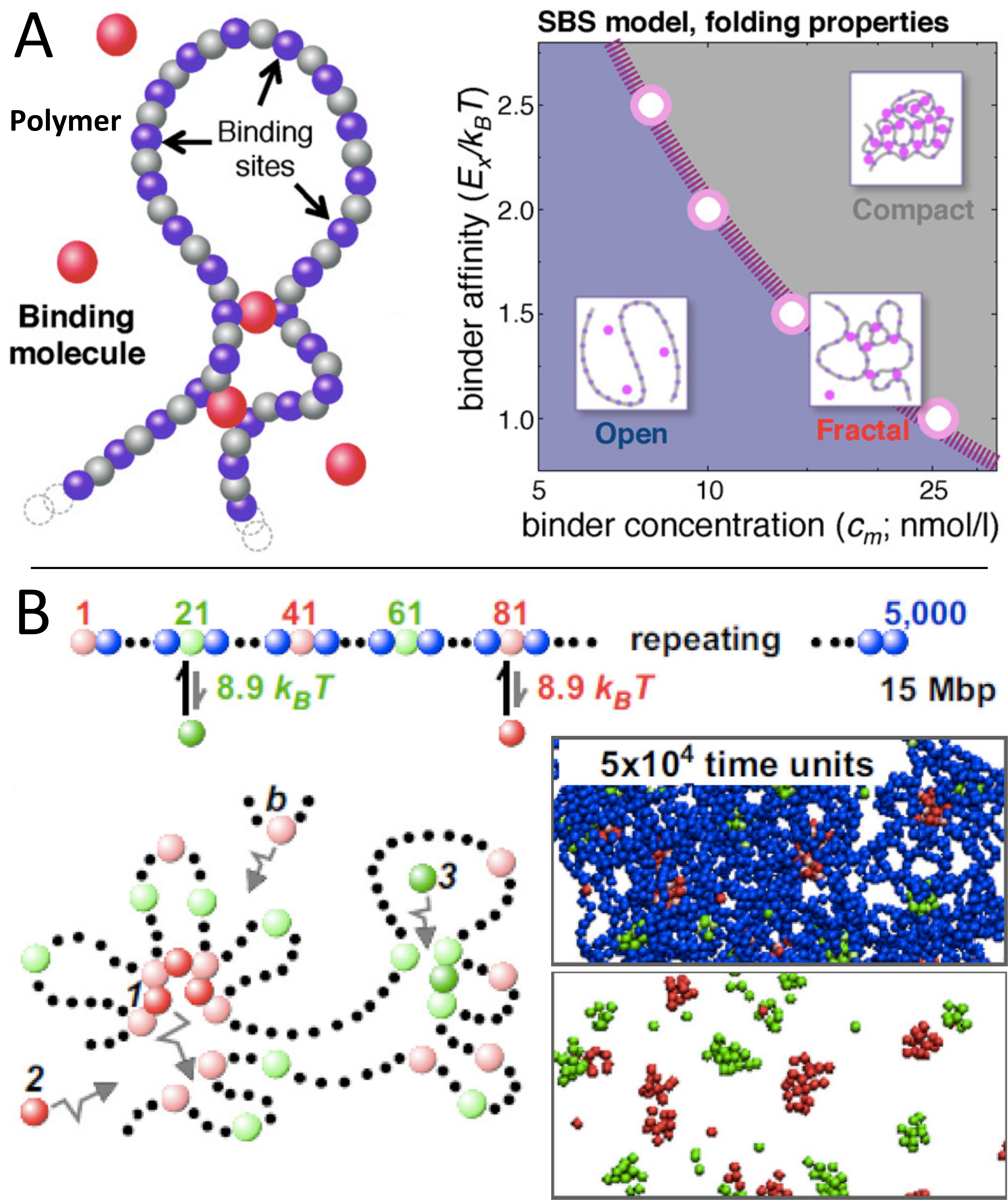}
    \caption{
    {\small
    The role of sequence-specific chromatin-chromatin interactions in chromosome organization. I.
    (A)
    In the ``strings-and-binders-switch'' (SBS) model, chromatin acts as a block copolymer with site-selective affinity $E_X$ for specific binders at concentration $c_m$.
    Chromatin folding/unfolding 
    can be represented 
    in terms of the phase diagram in these two parameters. 
    Reproduced with permission from Ref.~\cite{BarbieriPomboNicodemiPNAS2012}.
    (B)
    Protein-like particles mimicking transcription factors binding to cognate sites on a block copolymer model promote chromosome compaction by forming rosettes and TAD-like domains.
    The model predicts also the spontaneous self-assembly of proteins into factories.
    Reproduced from Ref.~\cite{BrackleyMarenduzzoNAR2016} under Creative Commons CC BY License.
    }
    }
    \label{fig:ChromosomesPolymerModels-pt1}
\end{figure}

In the so-called ``strings-and-binders-switch'' (SBS) polymer model~\cite{BarbieriPomboNicodemiPNAS2012},
chromatin is described as a {\it block copolymer} where a certain fraction of monomers (the ``binders'') act as binding sites for freely diffusive particles, see Fig.~\ref{fig:ChromosomesPolymerModels-pt1}(A).
The binding of particles to DNA is dynamic (binders attach and detach intermittently at finite rates), the mechanism being described in terms of two phenomenological parameters:
the binder affinity ($E_X$)
and
the binder concentration ($c_m$).
It is then possible to construct a phase diagram in the $E_X$-$c_m$ space where a single line separates swollen from compact polymer conformations, as in the classical $\theta$-collapse~\cite{RubinsteinColbyBook,BarbieriPomboNicodemiPNAS2012} in polymer physics.
The SBS model predicts that as per adaptation to continuously-changing external conditions chromatin is switching between these two states through a suitable combination of the concentration/affity of the binders,
thus accounting qualitatively for the observed fluctuations in chromatin loci spatial positions and contacts as measured in FISH and HiC.

In a variation of the SBS-model, Brackley {\it et al.}~\cite{BrackleyMarenduzzoNAR2016} pointed out that protein-like particles mimicking transcription factors which bind reversibly to cognate sites on a block copolymer model promote chromosome compaction, see Fig.~\ref{fig:ChromosomesPolymerModels-pt1}(B).
This model outline a picture where a single chromosome is organized into spatial motifs like rosettes and topological domains similar to the ones observed in HiC experiments.
Interestingly, as a by-product the model predicts that proteins self-organize into clusters (or, factories~\cite{CookTF2010}) due to a ``bridging-induced attraction'' which is mediated by polymer folding.

\begin{figure}
    \centering
    \includegraphics[width=0.45\textwidth]{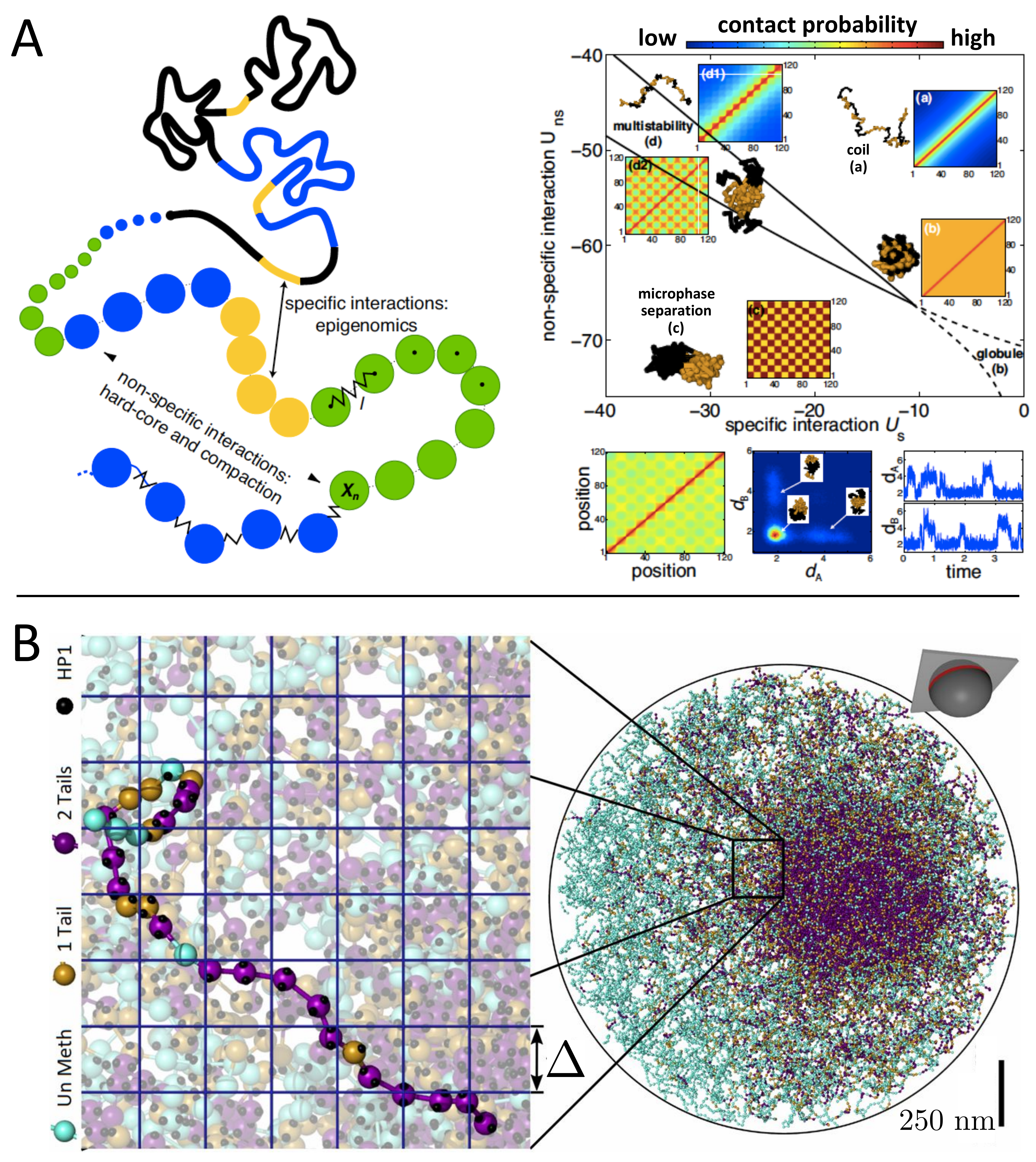}
    \caption{
    {\small
    The role of sequence-specific chromatin-chromatin interactions in chromosome organization. II.
    (A)
    Chromosomes may fold due to epigenome-specific attractive interactions promoting microphase segregation and TAD-like organization.
    Reproduced from Ref.~\cite{JostVaillantNAR2014} under Creative Commons CC BY License.
    (B)
    Trimethylated chromosomal sites attracting each other by the mediated action of oligomerized HP1 model proteins drive phase segregation into compact, heterochromatin domains {\it vs.} swollen, euchromatin domains.
    Reproduced with permission from Ref.~\cite{MacPhersonSpakowitzPNAS2018}.
    }
    }
    \label{fig:ChromosomesPolymerModels-pt2}
\end{figure}

Alternatively (or, in addition) to the action of the binders, the observed chromosome organization may be the consequence of the partitioning into a small~\cite{SextonCavalliCell2012} set of distinct epigenomic domains which cluster together by {\it epigenome-dependent} attractive interactions.
Jost and collaborators~\cite{JostVaillantNAR2014,GhoshJostPlosCB2018} have implemented this idea into a copolymer model, where each monomer of a specific epigenomic domain bind exclusively to monomers of the same species. The chromatin fiber associated to each chromosome thus segregate by a physical mechanism known as {\it microphase separation} (see Fig.~\ref{fig:ChromosomesPolymerModels-pt2}(A)) which displays a checkerboard pattern of contacts which may explain chromosome structure into TADs (reported in Sec.~\ref{sec:Intro}).
In a related study involving a very similar computational set-up, Shi {\it et al.} have shown~\cite{ShiHyeonThirumalaiNatComm2018} that chromatin dynamics is highly heterogeneous, reflecting the observed cell-to-cell variations in the contact maps:
folding is a two-step, hierarchical process which involves the formation of TAD-like chromatin domains (or, droplets) followed by their ``fusion'' inside the entire territory.

An interesting hypothesis on the connection between epigenetic marks (specifically, histone methylation) and chromosome folding has been recently formulated by MacPherson {\it et al.}~\cite{MacPhersonSpakowitzPNAS2018}.
By using Monte Carlo computer simulation of a nucleosome-resolved polymer model complemented by H3K9me3-methylation patterns from ChIP-seq data, the authors suggested that dimerization of HP1 single protein units which bind preferentially to methylated chromatin sites drive chromatin segregation into heterochromatin (dense and H3K9me3-rich) and euchromatin (open and H3K9me3-poor) domains, see Fig.~\ref{fig:ChromosomesPolymerModels-pt2}(B).
The segregation results in plaid-patterned heat-maps resembling those obtained in HiC experiments.

\subsubsection{III. Out-of-equilibrium effects: loop extrusion and activity-induced phase separation}\label{sec:ActiveLoopExtrusion}
\begin{figure*}
    \centering
    \includegraphics[width=1.0\textwidth]{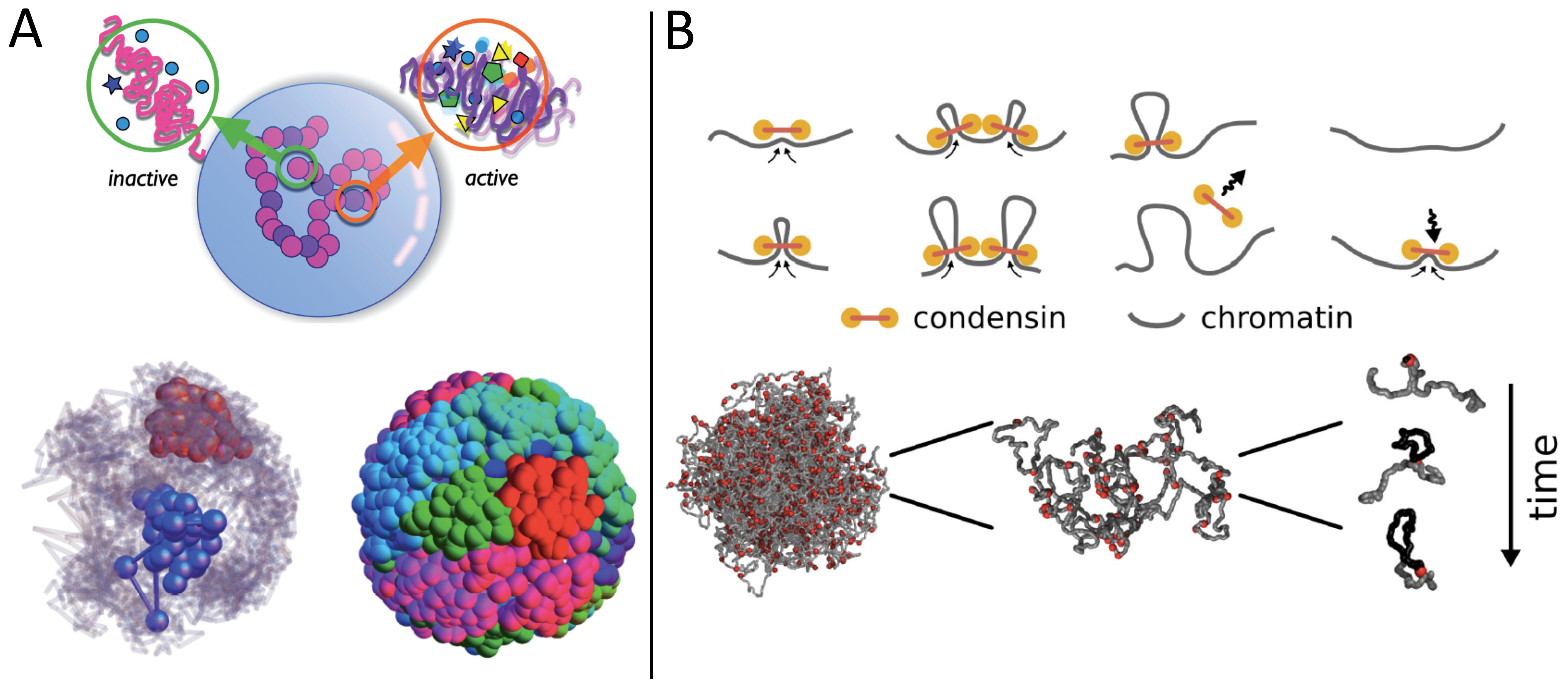}
    \caption{
    {\small
    The role of active processes in chromosome organization.
    (A)
    Chromatin is classified as ``inactive'' and ``active'' depending on its gene content (top).
    Gene-poor and gene-rich chromosomes phase separate and form territories whose spatial positions with respect to the nucleus correlate with experimental observations (bottom).
    Reproduced from Ref.~\cite{GanaiMenonNAR2014} under Creative Commons CC BY License.
    (B)
    The condensin complex (in yellow) bind to the chromatin fiber (in black) and, by moving into opposite directions, effectively produces chromatin loop extrusion.
    Extrusion stops when two (or more) complexes bump unto each other (top).
    An apparently disordered tangled mass of chromatin can then self-organize into a regular array of extruded loops (bottom).
    Reproduced from Ref.~\cite{GoloborodkoMirnyMitosisLoopExtrusion2016} under Creative Commons Attribution License.
    }
    }
    \label{fig:ChrsActiveProcesses}
\end{figure*}

Life is a dynamic process maintained through the continuous contribution of external energy sources:
as such, in recent years a conspicuous body of work on the experimental and theoretical aspects of non-equilibrium physics has had a tremendous impact on our understanding of how living matter works~\cite{RamaswamyAnnRevCMPhys2010}.
In this respect, chromosomes are no exception.
In the following, we summarize a few works which have contributed to highlight the role of non-equilibrium mechanisms with regard to chromosome organization.

Ganai {\it et al.}~\cite{GanaiMenonNAR2014} suggested that certain reported correlation between chromosome positioning within the nucleus and gene density (see Sec.~\ref{sec:Intro}) can be understood as the consequence of different ``activity'' levels:
similarly to the approaches described in previous sections
chromosomes are modeled as coarse polymers,
however -- in contrast to the purely passive systems discussed so far -- here each monomer is classified according to its level of activity (proportional to gene density) and coupled to a specific, effective temperature.
Thus, a higher effective temperature means a larger activity.
With the addition of a given amount of permanent loops between chromatin fibers, this models shows that chromosomes tend to be partitioned into clusters of different temperatures, see Fig.~\ref{fig:ChrsActiveProcesses}(A).
A rigorous physical explanation of this phenomenon was provided in Ref.~\cite{GrosbergJoanny2015} and later confirmed in Ref.~\cite{SmrekKremerPRL2017} by means of systematic computer simulations:
even small temperature gaps induce phase separation in systems of colloids or polymer chains.
In spite of the intrinsic {\it out-of-equilibrium} nature of the system,
it can nonetheless be shown that the phenomenon can be captured by the analogy to the classical {\it equilibrium} theory of binary mixtures which phase separate as the result of distinct chemical affinities~\cite{RubinsteinColbyBook}.

Recently, it has been pointed out that {\it active loop extrusion} may be universally responsible for chromosome segregation during mitosis~\cite{GoloborodkoMirnyMitosisLoopExtrusion2016,GoloborodkoMarkoMirnyBJ2016} and for chromosome compartmentalization into TADs~\cite{FudenbergMirnyCellRep2016}.
Specific proteins called ``condensins'' assemble into complexes and bond together spatially close loci on the chromatin fiber, see Fig.~\ref{fig:ChrsActiveProcesses}(B).
Then, the chromatin filament fixed by the condensins starts to be effectively {\it extruded} when the complex moves into opposite directions along the fiber.
When two condensins collide into each other the translocation process stops.
Moreover, with the addition of topoisomerase-II the loop extrusion mechanism is able to simplify chromosome topology by removing knots and links~\cite{RackoStasiakPolymers2018,OrlandiniMarenduzzoMichielettoPNAS2019} between chromatin fibers within the crowded environment of the nucleus.


%
\subsection{Building chromosomes by data-driven, ``top-down'' polymer models}\label{sec:TopDownModels}
The polymer models illustrated in Sec.~\ref{sec:BottomUpModels} employ minimal physical assumptions in trying to capture various aspects of chromosomes phenomenology
and, for this reason, they have been generically termed ``bottom-up''.
The most fascinating side of these approaches is that they often make testable predictions which are amenable to experimental validation.

Recently, a number of studies have attacked the problem of chromosome organization from a radically different perspective:
instead of explaining experimental observation by employing minimal physics why not using the information contained in the experiments to deduce the {\it most probable} chromosome conformations compatible with the observations?

\begin{figure*}
    \centering
    \includegraphics[width=0.95\textwidth]{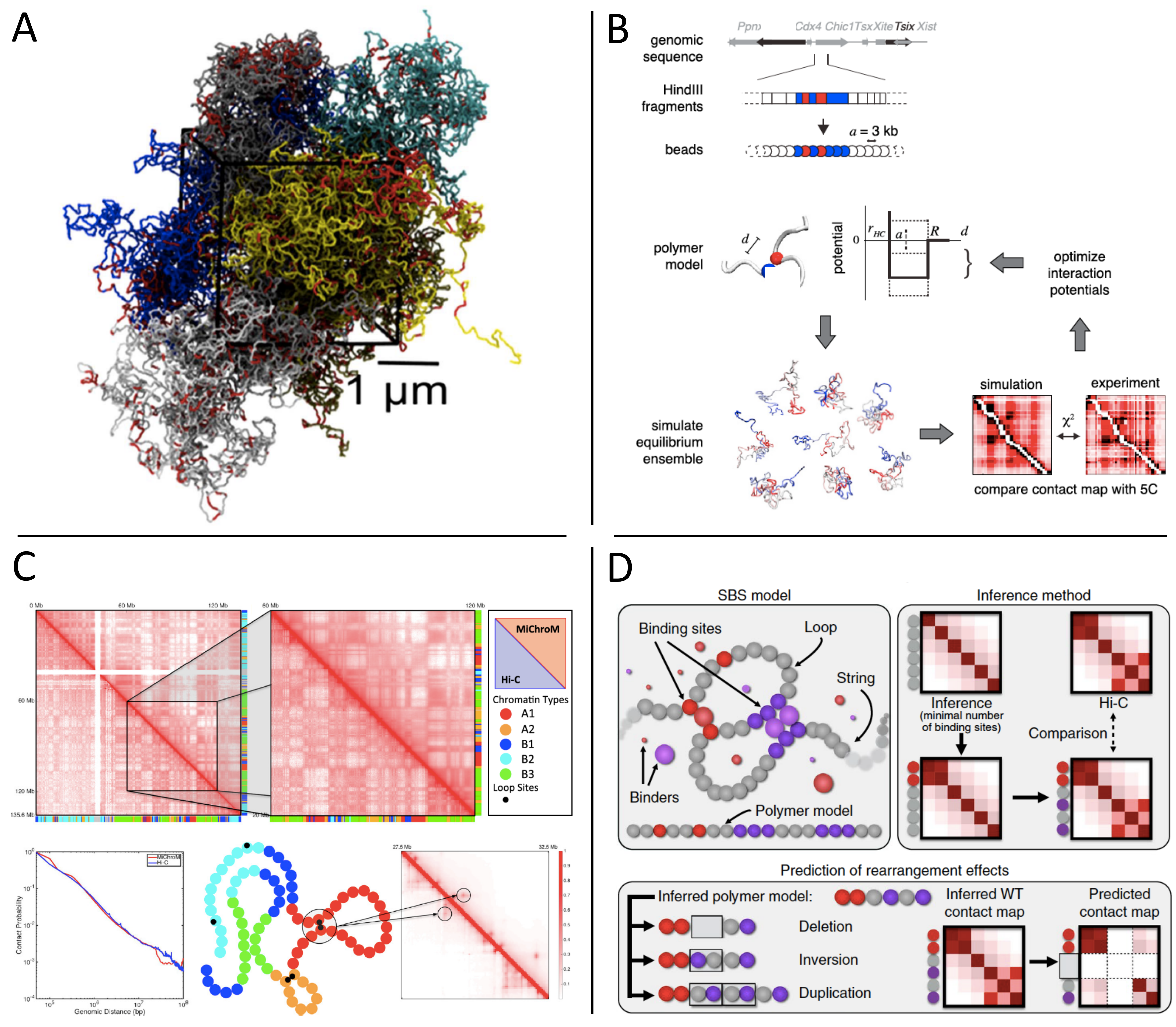}
    \caption{
    {\small
    Data-driven polymer models.
    (A)
    A polymer model promoting colocalization of coexpressed genes in human chromosome 19 produces conformations organized in spatial macrodomains which correlate with HiC~\cite{LiebermanAiden-HiC-Science2009} predictions.
    Reproduced from Ref.~\cite{DiStefanoRosaMichelettiPlosCB2013} under Creative Commons License.
    (B)
    A single TAD is modeled as a polymer chain whose beads interact {\it via} a square-well potential with an attractive wall. The energy parameters are optimized by iteration of a Monte Carlo sampling scheme so to maximize the agreement between the {\it predicted} and {\it observed} chromosome conformation contacts maps.
    Reproduced with permission from Ref.~\cite{GiorgettiTianaHeardCell2014}.
    (C)
    In the Minimal Chromatin Model (MiChroM), chromatin loci are classified into different types (colors) and certain pairs of genomic loci (``anchors'') tend to form loops.
    The interaction potential for the polymer chain is {\it trained} based on the HiC~\cite{LiebermanAiden-HiC-Science2009} contact matrix for human chromosome 10,
    and used then to construct and study the spatial features of the other chromosomes.
    Reproduced with permission from Ref.~\cite{DiPierroOnuchicPNAS2016}.
    (D)
    The Polymer-based Recursive Statistical Inference Method (PRISMR) refines the SBS polymer model~\cite{BarbieriPomboNicodemiPNAS2012} by ``filtering'' the simulated chromosome conformations so to derive the minimal set of binding sites and binding molecules which best reproduces the input HiC contact matrix.
    Instructing the model on wild-type (WT) chromosome data, the effects of genomic mutations (deletions/inversions/duplications) on abnormal chromosome conformations can be then predicted without further additional parameters.
    Reproduced with permission from Ref.~\cite{BiancoPomboNicodemi2018}.
    }
    }
    \label{fig:DataDrivenModels}
\end{figure*}

In two related studies, Di Stefano and coworkers showed that by just enforcing co{\it localization} of co{\it expressed} genes in a polymer model for human chromosome 19 first~\cite{DiStefanoRosaMichelettiPlosCB2013} and then for the entire human genome~\cite{DiStefanoMichelettiSciRep2016}
without major additional constraints,
the resulting conformations (see the example shown in Fig.~\ref{fig:DataDrivenModels}(A)) appear compatible with chromatin classification in A/B sub-domains
and with the non-random locations of chromosome territories correlated to gene content (see Sec.~\ref{sec:Intro}).

In order to exploit the nature of TADs and of chromatin-chromatin interactions measured within a single TAD, Giorgetti {\it et al.}~\cite{GiorgettiTianaHeardCell2014} introduced a computational polymer model (see. Fig.~\ref{fig:DataDrivenModels}(B)) where sequence-dependent monomer-monomer interactions were obtained upon maximizing the agreement between contact frequencies predicted by the model and the ones measured by ordinary conformation capture techniques.
The model, targeted onto a specific region of mouse chromosome X, reveals that the structure of a single TAD measured by HiC reflects a full ensemble of fluctuating conformations across the cell population with no stable loops.
Interestingly, the model was later tested by inducing a deletion at a specific locus and measuring the altered spatial distances.

A similar approach, the Minimal Chromatin Model (MiChroM), was introduced recently by Di Pierro {\it et al.}~\cite{DiPierroOnuchicPNAS2016} with the intent of expanding the analysis to an entire chromosome and trying to export the derived force-field to describe the whole diploid nucleus.
Specifically, polymer loci were classified into chromatin types (as in some of the models considered in Sec.~\ref{sec:InteractionsRole}) and the energy parameters describing the interactions between them were trained by using HiC data for human chromosome 10 from a specific cell line, see Fig.~\ref{fig:DataDrivenModels}(C).
The model was then used {\it to predict} an ensemble of possible structures for the other chromosomes not used for the training of the energy function:
interestingly, the obtained maps match well the ones obtained by HiC and the simulated chromosome structures recapitulate other notable features of interphase chromatin, like microphase separation of chromatin types (Sec~\ref{sec:InteractionsRole}) and the tendency of open chromatin to remain at the periphery of its territory.

Finally, Bianco {\it et al.} refined the SBS model discussed in Sec.~\ref{sec:InteractionsRole}, by introducing the Polymer-based Recursive Statistical Inference Method (PRISMR)~\cite{BiancoPomboNicodemi2018}:
PRISMR works by minimizing a cost function which -- again -- takes into account the predicted {\it vs.} the measured HiC contact frequencies, see Fig.~\ref{fig:DataDrivenModels}(D).
The ``optimal'' polymer model is then exported to construct chromosome conformations for a number of so-called structural variants of chromosomes which are known to produce anomalous chromatin folding and diseases.
The protocol is then shown to be very efficient in detecting mutated chromatin-chromatin interactions which are involved in anomalous phenotypes: the work reports in particular the example of the {\it EPHA4} locus where specific deletions are associated to anomalous polydactyly.

%
%


%
\section{Discussion}\label{sec:Disc}

In this article, we have described some of the most popular modelling approaches to 1D and 3D features of genomic DNA sequences.

With regard to 1D features, we have shown (Sec.~\ref{sec:1dModels})  evidence of nontrivial displacement of nucleotides along the sequence:
(1) at the single-nucleotide level, since pyrimidines and purines are not randomly distributed but show long-range correlations up to kb scale (Sec.~\ref{sec:RWsDNAseqs})
and
(2) at a dinucleotide level (Sec.~\ref{sec:DinuclInterdists}),
in particular CG-dinucleotides associated to DNA methylation, for which the distribution of mutual interdistances along the genome shows a different behaviour from the other dinucleotides and seems correlated to specific regulation mechanisms (CpG islands) or to organism complexity.
Thus, the analysis of 1D sequences in these specific cases reveals important properties that go {\it beyond} the 1D environment itself, and likely have an impact on (or are influenced by) the surrounding 3D context.

In order to understand how genomes fold in 3D we have presented recent work about molecular modeling (Sec.~\ref{sec:3dModelsChrOrg}) of chromosomes.
In this respect, the state-of-the-art is remarkably complex:
topological effects (Sec.~\ref{sec:TopologyRole}), specific DNA-DNA interactions (Sec.~\ref{sec:InteractionsRole}), energy-driven, active (opposed to entropy-driven, passive) mechanisms (Sec.~\ref{sec:ActiveLoopExtrusion}) are all likely to act concurrently.
Future work has to dissect one by one each of these mechanisms with the goal to understand their relative importance with respect to the full picture.

Inspired by the phenomenology of the ``protein folding'' problem~\cite{PandeGrosbergTanaka-RMP2000} where the aminoacid sequence  contains the essential information to drive the protein towards its unambiguous, ``native'' structure, it is natural to ask to which extent the 1D sequence influences the 3D chromatin architecture, provided that epigenetic factors are a key player to be associated to DNA sequence.
Two recent complementary approaches suggested that a significant amount of spatial contacts detected by chromosome conformation capture techniques can be predicted based on the spatial colocalization of transcription-factor binding sites measured by ChIA-PET~\cite{Szalaj-GenomRes2016} or from 1D maps of histone modifications and other epigenetic marks~\cite{Zhu-EpiTensor-NatureComm2016}.
However, in spite of some evidence pointing to some non-trivial interplay between 1D sequence and 3D folding, the full picture remains poorly understood.

In this respect, some recent attempts (Sec.~\ref{sec:TopDownModels}) based on ``data-driven'' polymer physics with input from epigenetic patterns seem to describe well the spatial structure of chromosomes {\it in vivo} and, in some specific cases, are able to identify critical hot-spots along the sequence associated to mutations in the phenotype.
At the same time, the 3D chromosome conformation participates actively in the occurrence of epigenetic phenomena along the 1D sequence, such as the formation of loops between specific chromatin loci having distant locations along the sequence.
Therefore, it appears plausible 
that the 3D 
chromosome organization 
is ``echoed'' in the positioning along the DNA sequence of 1D motifs associated to promoters and enhancers regulating gene expression~\cite{PomboDillonReview2015},
and that it is a major ``driving force'' in fixing and stabilizing the complex architectures~\cite{deJong2002,LagomarsinoRemondiniReview2009} of gene regulatory networks. 

Providing answers to these questions represents an exciting challenge which requires concerted experimental and theoretical efforts:
the hope of the future is to find a systematic way for addressing unsolved biological and medical challenges linking DNA sequences anomalies, chromosome misfolding and aberrant phenotypic behavior. 
A combination of 1D and 3D genome information can improve the understanding of pathologies with a ``structural'' basis, such as the Hutchinson-Gilford progeria syndrome in which a protein associated to nuclear membrane scaffolding and DNA arrangement is mutated~\cite{McCord13}, or of pathologies such as cancer~\cite{Rippe2019}, characterized by significant expression {\it de}regulation due to epigenetic phenomena and in which specific 1D mutational events can be associated to DNA 3D structure~\cite{Nicodemi19}.







\section{Acknowledgements}
AR and DR would like to acknowledge networking support by the COST Action CA18127.
DR and AM would like to acknowledge support by the  HARMONY IMI-2 n. 116026.


\bibliography{biblio}

%
\end{document}